\documentclass[sigconf, breaklinks=true]{acmart}
\newcommand{\an}[2]{#1} 

\usepackage{balance}
\usepackage[inline]{enumitem}
\usepackage{microtype}
\usepackage{multirow}
\usepackage{array}
\usepackage[utf8]{inputenc}

\copyrightyear{2025}
\acmYear{2025}
\setcopyright{rightsretained}
\acmConference[SIGCSE 2025]{Proceedings of the 56th ACM Technical Symposium on Computer Science Education V. 1}{February 26--March 1, 2025}{Pittsburgh, PA, USA}
\acmBooktitle{Proceedings of the 56th ACM Technical Symposium on Computer Science Education V. 1 (SIGCSE 2025), February 26--March 1, 2025, Pittsburgh, PA, USA}
\acmDOI{}
\acmISBN{}

\makeatletter
\gdef\@copyrightpermission{
  \begin{minipage}{0.3\columnwidth}
   \href{https://creativecommons.org/licenses/by/4.0/}{\includegraphics[width=0.90\textwidth]{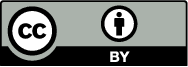}}
  \end{minipage}\hfill
  \begin{minipage}{0.7\columnwidth}
   \href{https://creativecommons.org/licenses/by/4.0/}{This work is licensed under a Creative Commons Attribution International 4.0 License.}
  \end{minipage}
  \vspace{5pt}
}
\makeatother

\begin{document}

\title[Code Interviews for Introductory Programming Assignments]{Code Interviews: Design and Evaluation of a More Authentic Assessment for Introductory Programming Assignments}

\author{\an{Suhas Kannam}{Anonymous}}
\orcid{\an{0009-0007-4620-5449}{}}
\affiliation{%
  \institution{\an{University of Washington}{Institution}}
  \city{\an{Seattle}{City}}
  \state{\an{WA}{State}}
  \country{\an{USA}{Country}}
}
\email{\an{ksuhas16@uw.edu}{email}}

\author{\an{Yuri Yang}{Anonymous}}
\orcid{\an{0009-0001-7740-5116}{}}
\affiliation{%
  \institution{\an{University of Washington}{Institution}}
  \city{\an{Seattle}{City}}
  \state{\an{WA}{State}}
  \country{\an{USA}{Country}}
}
\email{\an{yuriyang@uw.edu}{email}}

\author{\an{Aarya Dharm}{Anonymous}}
\orcid{\an{0009-0008-6905-6965}{}}
\affiliation{%
  \institution{\an{University of Washington}{Institution}}
  \city{\an{Seattle}{City}}
  \state{\an{WA}{State}}
  \country{\an{USA}{Country}}
}
\email{\an{adharm@uw.edu}{email}}

\author{\an{Kevin Lin}{Anonymous}}
\orcid{\an{0000-0001-9946-3635}{}}
\affiliation{%
  \institution{\an{University of Washington}{Institution}}
  \city{\an{Seattle}{City}}
  \state{\an{WA}{State}}
  \country{\an{USA}{Country}}
}
\email{\an{kevinl@cs.uw.edu}{email}}

\begin{abstract}
Generative artificial intelligence poses new challenges around assessment, increasingly driving introductory programming educators to employ invigilated exams. But exams do not afford more authentic programming experiences that involve planning, implementing, and debugging programs with computer interaction. In this experience report, we describe code interviews: a more authentic assessment method for take-home programming assignments. Through action research, we experimented with varying the number and type of questions as well as whether interviews were conducted individually or with groups of students. To scale the program, we converted most of our weekly teaching assistant (TA) sections to conduct code interviews on 5 major weekly take-home programming assignments. By triangulating data from 5 sources, we identified 4 themes. Code interviews (1) pushed students to discuss their work, motivating more nuanced but sometimes repetitive insights; (2) enabled peer learning, reducing stress in some ways but increasing stress in other ways; (3) scaled with TA-led sections, replacing familiar practice with an unfamiliar assessment; (4) focused on student contributions, limiting opportunities for TAs to give guidance and feedback. We conclude by discussing the different decisions about the design of code interviews with implications for student experience, academic integrity, and teaching workload.
\end{abstract}



\maketitle

\section{Introduction}

Generative artificial intelligence (AI) present new opportunities and new challenges for teaching, learning, and assessment in introductory programming courses \cite{10.1145/3568813.3600138, 10.1145/3623762.3633499, 10.1145/3545945.3569759, 10.1145/3568813.3600142, 10.5555/3636988.3637000}. The same capability that enables students to receive immediate, personalized, and helpful feedback to support their learning can also be used to circumvent learning. The availability and accessibility of generative AI poses a challenge to the accuracy of take-home assessment \cite{10.5555/3636988.3637000}. Educators ``need to prepare for a world in which there is an easy-to-use widely accessible technology that can be utilized by learners to collect passing scores, with no effort whatsoever, on what today counts as viable programming knowledge and skills assessments'' \cite{10.1145/3568813.3600142}.


While invigilated in-person pencil-and-paper exams remain a relatively reliable way to evaluate student learning outcomes, programming courses often include take-home work that resist time-limited and computer-free assessment. For example, educators may want to assess students' abilities to not only solve many small problems, but also assemble larger programs. This format might be harder to squeeze into small, fixed exam periods without computer access for students to run and debug code. Additionally, just as there are equity implications for the use of generative AI in computing education \cite{10.1145/3632620.3671116}, there are also equity implications for assessment policies as different decisions can impact student motivation, sense of belonging, and self-efficacy \cite{10.1145/3632620.3671100}. If authentic programming experiences are important learning objectives, how might we design a more robust assessment for take-home work while minimizing academic misconduct? How might we design more authentic and meaningful assessments that promote this goal of academic integrity while maintaining or increasing student motivation, sense of belonging, and self-efficacy? How might we achieve all of this without reinventing our existing courses and assignments?

In this experience report, we describe \textbf{code interviews}: a method to evaluate student proficiency on programming learning outcomes using live oral assessments about students' take-home work. We integrated our approach over two academic terms (quarters) in a large-enrollment undergraduate intermediate data programming course (an introductory python course taught in the context of data science). Drawing on action research methodology, we improved our approach over the two quarters using data triangulated from 5 different sources: observations, semi-structured focus groups, surveys, small-group instructional diagnosis, and teaching assistant (TA) feedback in order to answer three research questions.
\begin{enumerate}[label=\textbf{RQ\arabic*}, leftmargin=*]
    \item What experiences do students and TAs have with code interviews in an intermediate data programming course?
    \item What elements of code interviews do students find most beneficial in an intermediate data programming course?
    \item How can code interviews be improved for future intermediate data programming course offerings?
\end{enumerate}

Oral exams have long been used in education settings to assess student proficiency and learning outcomes \cite{10.1145/1151954.1067487}. In recent years, there have been many reasons for adopting oral exams in science, engineering, and computing courses, such as:
\begin{itemize}
    \item increasing confidence in assessments particularly due to disruptions of global health or new technologies \cite{10.1021/acs.jchemed.3c00011, 10.1145/3408877.3432511, 10.1109/FIE49875.2021.9637124};
    \item improving perceptions about assessment effectiveness \cite{10.1145/3408877.3432511, 10.1145/3287324.3287489, 10.1145/1151954.1067487} as well as motivation, stress level, overall course performance, and sense of belonging \cite{10.1145/3478431.3499382};
    \item emphasizing communication skills and assess student learning outcomes in light of generative AI \cite{10.5555/3636988.3637000}.
\end{itemize}
Results from these studies have been mixed. Depending on the approach taken, study population, and research methods, studies have varied in their ability to produce desirable educational outcomes. In undergraduate chemistry education, \citeauthor{10.1021/acs.jchemed.3c00011} report that ``Students had an overwhelmingly positive response to the oral exam experience and recommended their continued use in spite of [\textellipsis] the stress and anxiety of verbal presentation and the depth of understanding required to answer questions verbally'' \cite{10.1021/acs.jchemed.3c00011}. In undergraduate computing education, \citeauthor{10.1145/3408877.3432511} found that student experiences varied by self-identified demographic characteristics, with women, Black or African American, Hispanic or Latino, and some subgroups of first-generation college students reporting higher stress levels before oral exams than their majority group peers \cite{10.1145/3408877.3432511}. The authors also found similarly concerning patterns of inequity for overall course performance and sense of belonging. \citeauthor{10.5555/3636988.3637000} reiterate two problems about oral exams: ``(1) planning the timing and logistics and (2) evaluating students responses in an ethical and unbiased way'' \cite{10.5555/3636988.3637000}. Despite these challenges, many authors still consider oral exams a promising assessment \cite{10.1021/acs.jchemed.3c00011, 10.1145/3408877.3432511, 10.1109/FIE49875.2021.9637124, 10.1080/26939169.2021.1914527, 10.1145/3287324.3287489} with the potential to make assessment more authentic \cite{10.1080/26939169.2021.1914527} and promote academic integrity \cite{10.1109/FIE49875.2021.9637124}.

Our work aims to contribute to the field:
\begin{enumerate}
    \item An oral assessment format for take-home programming work inspired by software engineering but specific to our courses.
    \item Action research reflections on the design of our code interviews, which changed over time in response to preliminary student, TA, and researcher feedback.
    \item An evaluation of code interviews to answer the research questions with triangulation from several data sources.
\end{enumerate}

\section{Code Interviews}

Code interviews are a type of live oral assessment for programming. In the context of our course---an intermediate Python data science course at a large research-intensive university typically enrolling over 200 students---code interviews were offered only in-person and designed to evaluate students' ability to explain and extend their weekly take-home programming work. As this required discussing their take-home work, discussions were conducted on the due date for each of the 5 weekly programming assignments during the weekly 50-minute TA-led section. In order to scale code interviews, we over-staffed sections to a ratio of 3 TAs per 30 student section in order to host 3 five-minute code interviews in parallel during each section. This over-staffed section structure gave us space to experiment with research variables such as the number of students who conducted a code interviews at once (group size) and the type of questions TAs asked during code interview (and, therefore, the expected duration of each interview).

\subsection{Winter Quarter}

During winter quarter, code interviews were conducted in groups of 3 or 4, which enabled substantial conversation between students. Students discussed how differences in their approaches affected redundancy, maintainability, readability, and consistency, but they were not expected to answer specific questions. This often involved students showing their code to each other and explaining their problem-solving approach. Through listening and guiding this conversation, TAs would be able to evaluate learning outcomes by counting the number of substantial contributions that each student made during the group discussion: to receive full credit, each student needed to make 2 substantial contributions to the discussion.

All code interviews during winter quarter were conducted the same way. For each function, the TA could ask: Can someone explain how their solution to this problem works? How did you translate the problem requirements into a plan for coding? As students responded, the TA as well as other students were encouraged to ask follow-up questions, such as:
\begin{itemize}
    \item Can someone else explain how their implementation compares to this first approach?
    \item What are the advantages or disadvantages to your approach?
    \item Are there any parts of the code that are particularly difficult to understand, maintain, or modify as problem requirements change in the future?
    \item Are there any other approaches for this assignment?
\end{itemize}
Code interviews were evaluated by TAs using a discrete, 4-level rubric from greatest proficiency to least proficiency: Exemplary, Satisfactory, Approaching, and Unassessable.

Across the entire winter quarter, 76\% of all code interviews conducted received Exemplary marks while the remaining 24\% received Satisfactory marks. Assignment completion rates as well as code interview participation rates were very high as they were both required by the course.

\subsection{Spring Quarter}

In the following spring quarter, the course staff wanted to increase the ``signal-to-noise ratio'' of code interviews by asking more specific questions and assessing students in smaller groups (or even individually). Spring quarter code interviews focused on specific questions drawn from a selection of question types: tracing a test case, comparing approaches, identifying an alternative approach, finding a bug, and writing code to solve an isomorphic problem. Each week, the course staff announced the types of questions that would be asked. During the code interview, TAs would choose questions from a question pool and evaluate the student's live oral response to it. If, after a short while, the student seemed stuck, TAs were empowered to give each student one nudge question to help them make progress. Proficiency was determined considering both the correctness of the answer and effectiveness of the explanation.

Across the entire spring quarter, 30\% of all code interviews received Exemplary marks, 63\% received Satisfactory marks, and the remaining 7\% received Approaching marks. Assignment completion rates as well as code interview participation rates were very high as they were both required by the course.

\subsubsection{First Code Interview}

The first code interview during spring quarter extended the weekly take-home programming assignment that reviewed Python programming concepts. This was conducted in groups of 2 students to better assess individual students while maintaining the social and comparative discussion aspect from winter quarter code interviews. During this code interview, we asked three types of questions to each student: tracing a test case, comparing approaches, and identifying an alternative approach.

\emph{Tracing a test case} involved explaining how a particular input is processed by a program: the control flow, the exact values at the beginning, end, and any other significant section, etc.

\emph{Comparing approaches} was the most similar to the winter quarter code interviews. For this task, students looked at each others' code for a particular function on their take-home programming work and compare their approaches with the same four values emphasized in winter quarter. This question allowed students to see each others' solutions and think critically about whether it was a separate way to solve the problem or a logically equivalent solution.

\emph{Identifying an alternative approach} required the student to solve one of the problems from their take-home programming assignment in a different way. Scenarios included constraints such as solving the problem without any data structures, with a different data structure, etc. This type of question pushed students to think about different the advantages and disadvantages of different approach.

These question types provided excellent coverage of the learning outcomes, but we faced several difficulties. There were too many questions for the time frame we prepared, so TAs and students felt rushed. Student work also often converged on the same approaches, so comparison between approaches was limited. Finally, students found it difficult to develop a new approach in front of a TA. We adjusted the next code interview according to these experiences.

\subsubsection{Second Code Interview}

The second code interview extended the corresponding assignment on statistical data analysis. To alleviate time pressure, we moved from three questions to two questions with the same time limit. The two types of questions were tracing a test case (as before, intended to help warm up) and finding a bug.

\emph{Finding a bug} involved identifying the defects in a TA-provided sample solution to the weekly programming assignment. Depending on the particular question, the output of running the defective code was sometimes given as a nudge and sometimes given initially with the problem itself. This question evaluated students' debugging skills as well as their understanding of their own code. When the output was not given, this question tested students' understanding of the problem and code comprehension in order to to identify the approach and the potential bug within it.

For this code interview, we intentionally allowed some TAs to conduct code interviews individually while other conducted code interviews in groups of 2 students so that we could learn more about variations in group size. As questions became more structured and standardized, we hypothesized that there would be less benefit to conducting code interviews in pairs. The TAs who conducted the interviews individually thought they could more effectively assess a student's understanding of the concepts and ensure that no student's grade is riding on who their partner is for the interview. But the TAs who conducted the interviews in pairs felt students would not be as nervous or stressed for the interview. 

\subsubsection{Third Code Interview}

The third code interview extended the corresponding assignment on data visualization. This code interview was conducted entirely individually: TAs met with students one-on-one. TAs asked two types of questions: finding a bug and writing code to solve an isomorphic problem.

\emph{Isomorphic problems} are very similar to problems in the weekly take-home programming assignment---designed so that students only needed to modify a few small parts of their programming assignment in order to solve the new isomorphic problem. TAs evaluated students on not only their solution to the isomorphic problem but also their explanation of their work.

The main difficulty with this structure was that the isomorphic question took a long time since students had to read the entire question again and then develop a solution, including dealing with any errors along the way. Time again became an issue, particularly since these code interviews were conducted individually. We decided that the next time we asked an isomorphic question to students individually, it would be the only question we asked.

\subsubsection{Fourth Code Interview}

The fourth code interview extended the corresponding assignment on software engineering. Since the engineering task involved implementing a complicated algorithm, TAs asked two types of questions: tracing their solution and finding a bug. We decided to conduct this code interview in pairs in response to feedback from students that interviews were stressful. We appreciated asking different types of questions each week: as assignments changed from week to week, code interview questions also changed to reflect them.

\subsubsection{Fifth Code Interview}

The fifth and final code interview extended the corresponding assignment on geospatial data visualization. For this interview, we wanted to focus on writing code to solve an isomorphic problem individually, which (as decided earlier) would be the only type of question we asked. Drawing from the third code interview where we last asked students to write code to solve an isomorphic problem, we wanted to run this fifth and final code interview the same way. However, taking into account feedback from students that they appreciated having a partner, we ultimately conducted this code interview in pairs but with separate questions and limited collaboration. We hypothesized that students might be able to gain some comfort from working alongside another student even if collaboration opportunities were limited.


\section{Methods}

We evaluated code interviews using weekly student surveys, observations of code interviews, semi-structured student focus groups, a class-wide small-group instructional diagnosis, as well as a few individual TA interviews and group TA discussions. We identified key themes in the data using inductive thematic analysis. None of the research instruments listed above were required, so student participation in the research evaluation was entirely optional.

Students were asked to complete a pre-survey, a post-survey, and 4 weekly surveys in between each of the 5 take-home programming assignments. The pre-survey included questions about students' comfort and familiarity with code interviews, as well as their sense of its importance. The post-survey included questions about students' perception of code interviews: what skills they believed they gained, self-confidence in their code interview skills, and how code interviews have change their perception on computing. At the end of the post-survey, we also asked students about demographic information to determine if there were disparities in experiences between students with different social identities; however, in this study, we did not utilize this information. For the weekly survey, we asked students the same set of three questions each week:
\begin{itemize}
    \item What is something you think you could have said or done differently during [code interviews]?
    \item What aspects of the [code interviews] did you find most valuable or challenging?
    \item What would you like different about the course structure and why?
\end{itemize}

During section, we observed groups of students as they completed their code interviews. Then, immediately after a group of students completed their code interviews, we conducted follow-up semi-structured focus groups to debrief with students about their experiences. These semi-structured focus groups were recorded and later transcribed. The number of students involved in each focus group depended on the group size: a focus group could include as few as 1 student and as many as 4 students. Our main questions for students were:
\begin{itemize}
    \item Could you describe an experience during [the preceding code interview] that you were proud of?
    \item What are the challenges of [code interviews]?
    \item What are the benefits of [code interviews]?
\end{itemize}

Our student-focused data also included small-group instructional diagnosis provided by educational consultants at our institution during lecture. Small-group instructional diagnosis uses a consensus-based approach involving small groups of 4 to 6 students, followed by discussion and polling of opinions with the entire class. Results were then analyzed by the educational consultants. Both the semi-structured student focus groups and small-group instructional diagnosis were informed by preliminary results from the pre-survey and weekly surveys: during these activities, we sought to better understand student experiences that were typically only very briefly mentioned in the surveys.
 
We also conducted a few individual TA interviews and group TA discussions. These were much less structured than the student focus groups and took place around staff meetings, immediately following the conclusion of a TA-led section, or during quiet periods in office hours. Our main questions for TAs were:
\begin{itemize}
    \item What aspects of [code interviews] could be improved?
    \item What patterns have you noticed among students that make mistakes?
    \item What patterns have you noticed among students that perform well?
\end{itemize}

Informed by action research methodology, we continuously integrated preliminary results to improve course operations and further data collection. During winter quarter, this primarily occurred by sharing preliminary results with students as class-wide feedback intended to improve their experience with code interviews. Acting on the preliminary winter quarter results and desires of the course staff, we overhauled spring quarter code interviews to emphasize more specific and standardized learning outcomes. Preliminary results identified during spring quarter then influenced variables like group size and question types.

After the conclusion of spring quarter, the first two authors reviewed all the data to inductively generate primary themes together. Since our data were primarily qualitative and our research questions primarily exploratory, the final list of themes reflect not only the perspectives that appeared most frequently, but also less common themes with valuable insights and experiences. The first two authors collected quotes from each of the data sources to place them in their respective themes, some of which are noted in the following results. The quotes presented were selected to reflect some of the variance and nuance in the data and should not be taken as a representative sample of each theme.

Some themes, however, were not relevant to our research questions so we chose to exclude them from the results. The most common excluded theme regarded uncertainty around how code interviews would factor into final grades. Although this theme represents a valid and critical question for our particular integration of code interviews, this theme did not relate particular qualities of code interviews to the research questions.

\subsection{Student Demographics}

Across both quarter, the self-reported student demographics were similar. Approximately 65\% of students identified as Asian, 25\% as White, and around 4\% as each of Black or Hispanic/Latino. Students of all other races collectively accounted for less than 3\%. For gender representation, there was a slightly higher proportion of women compared to men, less than 1\% identified as non-binary, and a small remainder preferred not to say. About 40\% of the class were English Language Learners, and roughly 10\% reported having disabilities.

Introductory and intermediate programming courses offered at our university have large enrollments typically between 200 and 800 students. For early-career undergraduate students, code interviews may have been the most personalized interactions that they have had with computing course staff about their programming work.

\section{Results}


\subsection{Code interviews pushed students to discuss their work, motivating more nuanced but sometimes repetitive insights}

Students highlighted the value of code interviews in reinforcing and deepening programming  concepts. They found that discussing their code helped them see it from new perspectives. One student noted, ``You realize something else about your code that makes you look at it completely differently'' (winter focus group). Another student remarked, ``Code interviews help you understand the code better'' (spring small-group instructional diagnosis). Particularly during winter quarter, we asked students to think about maintainability, different test cases, etc. with the intention of encouraging students to reevaluate their work from a different perspective. Although we saw this theme occur between both winter and spring quarters, we suspect that the effect size depends greatly on the particular interview question types asked. Our experimentation during spring quarter explored several question types in order to better align question type with the specific content of each assignment.

Because students all completed the same assignment each week, code interviews sometimes felt repetitive. During winter quarter where the emphasis was on comparison between approaches, ``if students are only giving one or two solutions it sort of contributes to [not enough people engaging in the code interview]'' (winter focus group). Another student noted that “the answers of how you did the code will be more similar so it's repetitive” (spring focus group). In group code interviews, if two students give two unique responses, other students with similar responses might not answer the question at all---thinking their response would not contribute to the conversation. This can limit student engagement during group code interviews with implications for equity in student participation, particularly considering how group code interviews combine student participation with evaluation.

\subsection{Code interviews enabled peer learning, reducing stress in some ways but increasing stress in other ways}

Code interviews enabled students to learn from each other. One student said, ``When I ran [into] an issue---or some problem someone else ran into---they [peers] explained how they solved it'' (spring focus group). Because students are given time to think about how to answer questions, code interviews also provided an opportunity for them to discuss the assignment and problems that occurred when completing it, uncovering common mistakes and fixes. During the spring small-group instructional diagnosis, students mentioned that seeing how their peers solved problems helped them understand the concepts. Another student noted that ``it's effective if you have two people sharing their inputs and outputs. As long as it's constructive criticism, there [are] always ideas that you have never thought of unless someone else says it, so it helps to have another input'' (spring focus group). Our section observations corroborated this perspective. In addition to discussing their approaches during the code interview, students also had a chance to discuss their work with each other during the remainder of the scheduled section time while TAs were busy interviewing other groups.

Conducting code interviews with groups of students may reduce stress compared to individual, one-on-one code interviews with a TA. Some students particularly appreciated group code interviews because they could discuss their solutions with peers to improve the quality of their weekly programming assignment submission. Although the benefits of peer learning may have occurred most in group code interviews, even individual code interviews may have still enabled peer learning. During spring quarter sections, we observed students talking among themselves to help improve each other's solutions and prepare for code interviews even during weeks when code interviews were conducted individually. Alternating between group code interviews, individual code interviews, and collaborative section practice activities may provide a diversity of ways for students to teach and learn from each other.

On the other hand, some students preferred individual code interviews as they felt it better motivated them to understand the nuances of their code. One student mentioned, ``I think you have to know everything about your code. This one's going to be better because you're just by yourself'' (spring focus group). Another student compared individual code interviews against group code interviews, noting that during group code interviews, ``I pretend like I know what they are talking about. When it's individual, I know what I'm talking about and I feel more confident'' (spring focus group). Although group code interviews may provide greater opportunity for peer learning, the ultimately individual assessments of learning outcomes can lead to students to focus on delivering their own explanations over learning other students' approaches.

This can lead to tension within groups of students: one group noted how group code interviews felt ``like a competition'' (winter focus group). Questions were asked to all the students in a group at the same time, so students who answered faster had more opportunity to demonstrate learning outcomes by contributing new insights to the discussion. The design of interview questions and the interpersonal dynamics of student groups may impact participation.

\subsection{Code interviews scaled with TA-led sections, replacing familiar practice with an unfamiliar assessment}

Compared to other introductory programming courses at our institution that typically utilize section time for review and the occasional quiz, code interviews were new to both students and TAs. One group of students noted, "I found it challenging not knowing which question would be asked and not having a lot to go off of for the debugging [interview questions]" (spring small-group instructional diagnosis). Although this often improved as students became more familiar with the expectations for code interviews, some students still noted an ``overall disconnect of what [the course staff] were expecting from students and how they performed'' even after completing several code interviews.

Additionally, TAs and students both identified time limitations as an issue, which was addressed through action research during both quarters but particularly during spring quarter with significant adjustments to the number of questions asked from 3 to 2 to 1 by the end of the quarter. TAs noted that language also posed a difficulty particularly for students whose first language was not English. One change inspired by this finding occurred during the spring quarter when course staff made refinements interview questions to emphasize the parts of isomorphic code writing problems that differed from the problems they completed at home.

Some students preferred review-based TA-led sections that are more common at our institution. One student suggested, ``It would also be helpful if the TA could write some code in section, like practice problems, to go over lecture concepts'' (winter focus group). Another noted, ``I like reviewing practice questions to understand what is taught during lecture'' (spring focus group). In the spring weekly survey, one student explained how going over practice question would ``improve understand[ing] of the concepts that are taught.'' By utilizing TA-led section time, we were able to overcome many scheduling and logistics challenges, but this came at the cost of removing familiar programming practice from the class. In the initial design for code interviews, this replacement was justified by a recent trend of abysmal student attendance rates in section, though future offerings may benefit from a more even balance between practice and assessment in section.

\subsection {Code interviews focused on student contributions, limiting opportunities for TAs to give guidance and feedback}

During winter quarter in particular, students were encouraged to lead discussions, but they often looked to TAs for structure and guidance. One student said, ``I do like when the teachers ask interjecting questions because they help us guide the conversation: `This is something that they---the teaching team---thinks would actually be valuable'{}'' (winter focus group). On the winter weekly survey, another student wrote, ``The most valuable part is that [my] TA will ask us for what kind of error we got most.'' More specific questions might elicit more insightful student contributions.

Not all students finished the weekly programming assignment by the time they conducted the code interview, and even students who finished might have benefited from specific TA feedback on writing high quality code. One student shared, ``I hope that the TA could help explain how to improve the code efficiency'' (spring focus group). Another student said, ``But I think it would be nicer if, during [code interview], they kind of give you a last minute, `Hey, make sure you're on track,' like, `This is the approach,' at least'' (spring focus groups). On the winter weekly survey one student wrote, ``I think that the TAs pointing out things that we might have missed helped because it allows me to check everything again with that in mind.'' The structure of code interviews as primarily an opportunity for students to demonstrate learning outcomes disempowered TAs from assisting students.

Although the course offers many office hours for students to receive personalized feedback, some students felt they would have benefited from feedback on their assignment code during or after the code interview. Code interviews may reveal new questions or insights, so there is an opportunity for TAs to support students in exploring these threads to support curiosity-driven learning.

\section{Conclusion}


As a primarily qualitative and exploratory experience report, our work raises questions for further study. Code interviews during both quarters had TAs asking questions and students answering them, but student and TA preferences about the specific structure and format were mixed. By spring quarter, we felt the most suitable structure and format depended on the particular assessments and particular learning objectives. But future work could envision code interviews as a more collaborative effort between students and TAs, blending elements of learning with assessment. For instance, allowing students to pose questions to peers or TAs could foster critical thinking and a deeper engagement with their code. Code interviews could be perceived as a more authentic task if students were given more space to engage their curiosity, creativity, collaboration, self-expression, and ways of knowing.

Additionally, our approach required triple-staffing sections with TAs, which is not sustainable in the long term without significant adjustments to TA responsibilities. While this workload is likely lower than the alternative of offering (and then grading) a paper-and-pencil exam during the same time frame, the high frequency of code interviews poses a challenge. Should frequent assessment be desirable, consider \emph{peer code interviews} whereby students interview each other. Could these peer code interviews enable peer learning, reduce student stress, and enable frequent assessment at scale without sacrificing accuracy? Future research could build on the literature around near-peer mentoring to guide the design of effective peer-led code interviews, exploring how TAs might best moderate this process and prepare students to act as interviewers.

During both winter and spring, code interviews supplemented traditional assessment of take-home programming assignments. However, code interviews could take a more central role in the assessment process, particularly as generative AI affects the importance of particular programming skills. In this context, code interviews have the potential to adapt to new demands by enabling authentic assessment of programming skills in a way that supports academic integrity. During spring quarter, code interviews focused on specific skills and learning outcomes that course staff prioritized, allowing TAs to observe students’ problem-solving processes and programming fluency more closely. Anecdotally, isomorphic code writing problems in particular provided TAs a very close look at students' problem solving process and programming fluency. 

While our results primarily focused on code interview formats, it is important to consider student demographics when designing assessments and course policies. About 40\% of students in both quarters identified as English Language Learners, which could have influenced perceptions about the stressfulness of the format. In addition to arriving at an answer to the TA's questions, students were also expected to clearly explain their answers too. Future work should explore the role of language as a mediating factor for code interviews. Additionally, the particular TA--student pairings could have implications for student comfort and performance. Are students who are interviewed by TAs who share the same social identity presentation as them more inclined to feel more comfortable? Although representation among course staff and among students may appear similar at a macro level, the specific pairings within large classes may exacerbate inequities.

Finally, future work could also explore the potential for code interviews to serve as more diagnostic or formative assessments by supporting student learning in real time and encouraging reflection on coding strategies. Moving forward, code interviews may valuable not only as grading instruments but also as tools for directly supporting student learning, offering detailed feedback to students about how they might improve their programming process.

\section*{Acknowledgments}

\an{Ken Yasuhara for conducting the small-group instructional diagnosis. Our teaching teams during Winter 2024 and Spring 2024, especially Elizabeth Bui, Jainaba Jawara, Kai Nylund, Vatsal Chandel, Iris Zhou, and Arona Cho for contributions to the code interview infrastructure. Members of the Center for Learning, Computing, and Imagination at the University of Washington for feedback.}{Anonymized.}

\bibliographystyle{ACM-Reference-Format}
\balance
\bibliography{ms}

\end{document}